\documentclass[pre,aps,showpacs]{revtex4}
\newcommand{\beq}{\begin{equation}}
\newcommand{\eeq}{\end{equation}}
\newcommand{\bqn}{\begin{eqnarray}}
\newcommand{\eqn}{\end{eqnarray}}
\newcommand{\bqns}{\begin{eqnarray*}}
\newcommand{\eqns}{\end{eqnarray*}}
\newcommand{\bary}{\begin{array}}
\newcommand{\eary}{\end{array}}
\newcommand{\non}{\nonumber}

\usepackage{graphicx}%
\usepackage{amsmath}
\begin{document}
\baselineskip=20pt
\title{Power loss and electromagnetic energy density in a dispersive metamaterial medium}
\author{Pi-Gang Luan}
\address{Wave Engineering Laboratory, Department of Optics and
Photonics, National Central University, Jhungli 320, Taiwan}

\begin{abstract}%
The power loss and electromagnetic energy density of a
metamaterial consisting of arrays of wires and split-ring
resonators (SRRs) are investigated. We show that a field energy
density formula can be derived consistently from both the
electrodynamic (ED) approach and the equivalent circuit (EC)
approach. The derivations are based on the knowledge of the
dynamical equations of the electric and magnetic dipoles in the
medium and the correct form of the power loss. We discuss the role
of power loss in determining the form of energy density and
explain why the power loss should be identified first in the ED
derivation. When the power loss is negligible and the field is
harmonic, our energy density formula reduces to the result of
Landau's classical formula. For the general case with finite power
loss, our investigation resolves the apparent contradiction
between the previous results derived by the EC and ED approaches.
\end{abstract}
 \maketitle
%----------------------------------------------------------------------
\section{Introduction}

Artificial electromagnetic media having negative permittivity and
permeability have been fabricated and tested experimentally for
several years \cite{Review}. According to Veselago
\cite{Veselago}, these metamaterial media are left-handed (over a
finite range of frequency), in the sense that the Poynting vector
and wave vector are antiparallel to each other. Besides, they are
dispersive and absorptive in general \cite{Shelby}.

For a dispersive medium with negligible absorption, the energy
density formula can be obtained via analyzing an adiabatic
electromagnetic process \cite{Brillouin,Landau}. However, this
analysis does not work when finite absorption is present. To
evaluate the electromagnetic energy density stored in a dispersive
medium with nonzero absorption, one has to adopt different
strategies. Now the existence of the (effective) left-handed
metamaterial makes this problem even more dramatic because
negative permittivity and permeability seems to imply the
possibility of negative energy density, contradicting the
thermodynamic stability conditions.

If the absorption of the medium is infinitesimal, the
time-averaged energy density of a harmonic electromagnetic wave
would be given by \cite{Landau}: \beq
\langle{W\rangle}=\frac{\epsilon_0}{4}\frac{\partial(\omega\epsilon(\omega))}
{\partial\omega} |{\bf
E}|^2+\frac{\mu_0}{4}\frac{\partial(\omega\mu(\omega))}{\partial\omega}
|{\bf H}|^2,\label{classical}\eeq where ${\bf E}$ and ${\bf H}$
are the complex electric and magnetic fields, and
$\epsilon(\omega)$ and $\mu(\omega)$ denote the frequency
dependent permittivity and permeability, respectively. Hereafter
we name Eq.(1) as Landau's classical formula. This formula
provides a reference for checking the correctness of the desired
energy density formula in the lossless limit.

There are two common approaches, namely the equivalent circuit
(EC) approach and the electrodynamic (ED) approach, being used to
derive the energy density formula for a dispersive media with
finite power loss. In the EC approach, firstly one has to
transform the wave medium problem to a corresponding electric
circuit problem \cite{Tretyakov,Fung}, where the values of the
capacitances, inductances, resistances and their arrangements in
the circuit system can be deduced from the specific forms of
$\epsilon(\omega)$ and $\mu(\omega)$, and then the electric and
magnetic energies stored in the circuit system can be evaluated.
In the final stage, one transforms the result back to the original
wave medium problem to find the corresponding energy density. On
the other hand, in the ED approach, the energy density formula is
obtained as a byproduct of the following energy conservation law
(the Poynting theorem) \beq -\nabla\cdot{\bf S}=\frac{\partial
W}{\partial t}+P_{loss}.\label{conservation} \eeq This
conservation law can be derived using Maxwell's equations, with
the aid of the equations of motions of the polarization and
magnetization of the medium \cite{Loudon,Ruppin,Kong,Boardman}.
Here ${\bf S}$, $W$, and $P_{loss}$ stands for the Poynting
vector, energy density, and power loss, respectively. Usually the
EC approach provides the time-averaged result although the energy
density at a specific time can also be deduced. On the other hand,
the ED approach is inherently a time domain approach, which
provides the expression of the instantaneous energy density of an
arbitrarily varying electromagnetic field. The time-averaged
result for a harmonic wave can also be obtained by averaging the
energy density in one period of oscillation.

It has been pointed out, if the medium has finite power loss, it
is impossible to define the energy density uniquely if we do not
have a microstructure model of the material \cite{Tretyakov}. With
the microscopic models of the electric and magnetic constituents
of the medium, the dynamical behaviors of the corresponding
electric and magnetic dipoles can be predicted, and the energy
stored in the the medium can be correctly evaluated. In the
literature, several dispersive media with different microscopic
dipole models have been considered. The simplest one is an
absorptive classical dielectric (Lorentz dispersion) with a single
resonant frequency \cite{Loudon,Fung}. This can be generalized to
the case that both the permittivity and the permeability have
Lorentz type dispersions \cite{Ruppin,Kong}. Non-Lorentz type
dispersions have also been considered. For example, in the
wire-SRR metamaterial medium, the wires provide the plasma-like
dispersion for permittivity \beq
\epsilon(\omega)=1-\frac{\omega^2_p}{\omega\left(\omega+i\nu\right)},\label{epsilon}\eeq
whereas the SRRs (split-ring resonators) provide a non-Lorentz
type dispersion for permeability \beq
\mu(\omega)=1+\frac{F\omega^2}{\omega^2_0-\omega^2-i\omega\gamma}.\label{mu}\eeq
Here the parameters $\nu$ and $\gamma$ represent the absorption
effect of the wires and SRRs, and $F$ is a dimensionless factor.
Besides, $\omega_p$ and $\omega_0$ are the effective plasma
frequency of the wire medium and the resonant frequency of the SRR
medium, respectively.

Recently, two different expressions for the electromagnetic energy
density of the wire-SRR medium were derived using EC
\cite{Tretyakov} and ED \cite{Boardman} approaches. Although the
electric energy densities obtained in these two papers are
consistent, their results for the magnetic part are different. In
addition, the magnetic energy density formula of \cite{Tretyakov}
(Eq.(31)) does not reduce to the classical result \cite{Landau} in
the lossless limit. We find that this was caused by the fact that
in evaluating the total energy (Eq.(27)), the magnetic energy (of
form $1/2\,{\rm Re}({\cal M}I_0I^*)$) stored in the mutual
inductance ${\cal M}$ between two sub-circuits was not taken into
account by the author. On the other hand, the formula in
\cite{Boardman} (Eq.(19)) does reduce to the classical result in
the zero absorption limit. However, when we transform every term
in this formula to the corresponding EC system to find its
counterpart, an unphysical term $V^2_R/2C$ (originated from the
$\omega^2\gamma^2$ term in the numerator) appears. Here $C$ is the
capacitance in the RLC subcircuit of the EC, and $V_R$ is the
voltage difference between the two terminals of the resistance $R$
in the subcircuit (see Fig.2 of \cite{Tretyakov}).

In addition to the above mentioned problems, we also noted that in
general the derivation via ED approach does not provide a unique
answer. This is caused by the fact that up to now there is no
unique way to determine whether a term with the dimension of power
should be included in the time derivative of the energy density
$\partial W/\partial t$ or in the power loss $P_{loss}$. In fact,
if one does not know the correct form of the power loss, one can
always redefine the energy density and power loss as $W'=W+U$ and
$P'_{loss}=P_{loss}-\partial U/\partial t$, where $U$ is an
arbitrary bilinear function of ${\bf E}$ and ${\bf H}$. For
harmonic ${\bf E}$ and ${\bf H}$ fields, this modification does
not change the time-averaged value of $P_{loss}$(i.e.,$\langle
P'_{loss}\rangle=\langle P_{loss}\rangle$), because $\langle
\partial U/\partial t\rangle=1/T\int^T_0 \partial U/\partial
t\,dt=(U(T)-U(0))/T=0$ ($T$ is the period of oscillation).
However, usually the time-averaged value of $W$ will be modified.
This observation explains why the time-averaged power loss
formulas obtained in \cite{Tretyakov} and \cite{Boardman} are the
same, but their time-averaged energy density formulas are
different. This observation also reveals that the expression of
the energy density is related to the power loss we choose. Note
that when we go to the lossless limit, the ambiguity discussed
here disappears, and a unique energy density formula can be
obtained. However, for the finite loss case, to identify the
energy density directly is difficult and we do not know any
practical method to avoid the above mentioned ambiguity, thus we
propose to identify the power loss first.

In order to resolve the contradictions between the EC and ED
approaches and derive a unique and physically reasonable energy
density formula, we adopt the following criteria. First, the
results derived by using different approaches must be the same.
Second, the formula must reduce to Landau's classical formula in
the zero absorption limit. Third, the origin and the expression
for the power loss must be carefully analyzed and identified
first.

In this paper, we will show that the unique energy density formula
can be obtained by using either the ED or EC approach. In
addition, the comparison between these two different derivations
helps us to clarify the meaning of each physical quantity
appearing in the energy density formula. The essential part in the
ED derivation is the correct form of the power loss, and we show
that it can be found by carefully analyzing the heat generating
mechanism in the medium. Our discussion and obtained results in
this paper resolve the apparent contradictions between the ED and
EC approaches and correct the calculation errors in other previous
papers. Although in this paper we consider only the wire-SRR
medium, the method is in fact not restricted by this case and can
be applied to other kinds of dispersive metamaterial media as
well.

This paper is organized as follows. In section II we derive the
energy density formula via ED approach. We argue that in this
derivation the correct form of power loss is essential for
obtaining the unique result we desire. In section III, we further
establish the ED-EC correspondence by constructing the EC system
for evaluating the magnetic energy stored in the SRR array. The
ED-EC correspondence further confirms the correctness of the
energy density and power loss formulas obtained by ED approach. In
section IV we present the conclusion of this paper.

\section{Power loss and ED approach}

Now we consider the metamaterial medium consisting of metallic
wires and split-ring resonators. Under the influence of external
electromagnetic field, the wires respond to the field as electric
dipoles, whereas the resonators play the role of magnetic dipoles.
After averaging the dynamical behavior of these elements, the
electromagnetic properties of the medium can be described by an
effective theory, having the following macroscopic quantities as
dynamical variables: ${\bf E}$, ${\bf D}$, ${\bf B}$, ${\bf H}$,
${\bf P}$, ${\bf M}$. They satisfy the following constituent
relations: \beq {\bf D}=\epsilon_0{\bf E}+{\bf P}\label{DE}, \eeq
\beq {\bf H}=\frac{\bf B}{\mu_0}-{\bf M}\label{BH}. \eeq The
dynamic equations for ${\bf P}$ and ${\bf M}$ are given by \beq
\ddot{\bf P}+\nu\dot{\bf P}=\epsilon_0\omega^2_p {\bf E}
\label{P}\eeq \beq \dot{\bf M}+\gamma{\bf M}+\omega^2_0\int{\bf
M}dt=-F \dot{\bf H}, \label{M}\eeq which can be derived by
analyzing the currents flowing in the wires and the SRRs under the
influence of the applying electromagnetic fields. The
displacements of the charges in the wires lead to the electric
dipoles, and the total electric dipole moment per unit volume
defines the polarization ${\bf P}$. Therefore, the dynamic
equation of ${\bf P}$ follows the form of the equation of motion
for the charges. On the other hand, a time varying magnetic field
parallel to the axes of the SRR arrays induces the oscillating
currents in these SRRs. Suppose the current in an SRR is $I$, and
the effective cross section area of it is $A$, then $m=IA$ is the
magnetic dipole moment of the SRR. The magnetization ${\bf M}$ is
then defined by the total magnetic dipole moment per unit volume.
The dynamical equation for the currents flowing in the SRRs can be
derived by using the Faraday's law,and the dynamic equation for
${\bf M}$ follows the same form. Note that the term on the right
hand side of Eq.(\ref{P}) is proportional to the electric field
${\bf E}$, whereas the corresponding term in Eq.(\ref{M}) is
proportional to the time derivative of the magnetic field. This
difference is caused by the fact that the electric dipoles are
induced by the electric driving field, but the magnetic dipoles in
this system can only be induced by the time varying magnetic
fluxes through the SRRs. For the details of the derivation,
readers may refer to Ref.\cite{Pendry1,Pendry2,Chen}. Using
Eq.(\ref{P}) and Eq.(\ref{M}), and assuming the monochromatic
condition, the permittivity of Eq.(\ref{epsilon}) and permeability
of Eq.(\ref{mu}) can be obtained according to the definitions:
$\epsilon(\omega)=D(\omega)/(\epsilon_0E(\omega))$,
$\mu(\omega)=B(\omega)/(\mu_0H(\omega))$.

Now we derive the energy conservation law of the form \beq
-\nabla\cdot\left({\bf E}\times{\bf H}\right)=\frac{\partial
W_e}{\partial t}+\frac{\partial W_b}{\partial t}+P_{loss} \eeq
from Maxwell's equations and the dynamical equations of ${\bf P}$
and ${\bf M}$ (Eq.(\ref{P}) and Eq.(\ref{M})). According to
Amper\'{e}'s law and Faraday's law, we have \bqn
&&-\nabla\cdot\left({\bf E}\times{\bf H}\right) ={\bf
E}\cdot\frac{\partial {\bf D}}{\partial t} +{\bf
H}\cdot\frac{\partial {\bf B}}{\partial t}\non\\
&&=\frac{\partial}{\partial t}\left(\frac{\epsilon_0 {\bf
E}^2}{2}\right)+{\bf E}\cdot\frac{\partial {\bf P}}{\partial
t}\non\\ &&\hspace{3mm}+ \frac{\partial}{\partial
t}\left(\frac{\mu_0 {\bf H}^2}{2}\right)+\mu_0{\bf
H}\cdot\frac{\partial {\bf M}}{\partial t}. \eqn

The electric energy density $W_e$ and magnetic energy density
$W_b$ can be obtained by integrating the ${\bf E}\cdot\partial{\bf
D}/\partial t$ and ${\bf H}\cdot\partial{\bf B}/\partial t$ terms,
respectively. The loss term $P_{loss}$ can also be obtained from
them. Note that the loss term cannot be written as a total
derivative, and this feature was utilized by the authors of
Ref.\cite{Boardman} to find the energy density. However, as we
have mentioned before, to uniquely determine the form of the
energy density, one has to carefully analyze the origin and the
correct form of the power loss first. Once the power loss has been
made certain, the energy density can be determined automatically.

The origin and form of the power loss in the wire-SRR medium can
be made certain by noticing the following two facts. First, both
$\dot{\bf P}$ and ${\bf M}$ are proportional to the currents
flowing in the conducting constituents (wires and SRRs) of the
wire-SRR medium. Second, the power loss of this medium can only be
originated from the Joule heat of form $I^2R$, generated in theses
conducting elements. We thus conclude that the power loss of the
wire-SRR medium should have the form \beq P_{loss}=\alpha \dot{\bf
P}^2+\beta {\bf M}^2,\label{powerform}\eeq where $\alpha$ and
$\beta$ are two appropriate constants.

Using Eq.(\ref{P}), we get \bqn {\bf E}\cdot\frac{\partial {\bf
P}}{\partial t}&=&\frac{1}{\omega^2_p\epsilon_0}
\left(\frac{\partial^2 {\bf P}}{\partial t^2}+\nu\frac{\partial
{\bf P}}{\partial t}\right)
\cdot\frac{\partial {\bf P}}{\partial t}\non\\
&=&\frac{\partial}{\partial t}\left(\frac{\dot{\bf
P}^2}{2\omega^2_p\epsilon_0}\right)+\frac{\nu}{\omega^2_p\epsilon_0}\dot{\bf
P}^2, \label{energyandloss}\eqn thus the electric energy density
$W_e$ should be defined as \beq W_e=\frac{\epsilon_0{\bf
E}^2}{2}+\frac{\dot{\bf
P}^2}{2\omega^2_p\epsilon_0}.\label{We}\eeq Note that the
additional term ${\nu}\dot{\bf P}^2/{\omega^2_p\epsilon_0}$ in
Eq.(\ref{energyandloss}) is the electric part of $P_{loss}$,
consistent with Eq.(\ref{powerform}).

The derivation of magnetic energy density is a little different,
as will be shown below. Substituting Eq.(\ref{M}) into the ${\bf
H}\cdot {\partial {\bf M}}/{\partial t}$ term, we have \bqn
&&\mu_0{\bf H}\cdot\frac{\partial {\bf M}}{\partial
t}=\frac{\partial }{\partial t}\left(\mu_0{\bf H}\cdot{\bf
M}\right)-\mu_0{\bf M}\cdot\frac{\partial{\bf
H}}{\partial t}\non\\
&&= \frac{\partial }{\partial t}\left(\mu_0{\bf H}\cdot{\bf
M}\right)+\frac{\mu_0}{F}{\bf M}\cdot\left(\dot{\bf M}+\gamma
M+\omega^2_0\int{\bf M}dt\right)\non\\
&&=\frac{\partial }{\partial t}\left[\mu_0{\bf H}\cdot{\bf M}+
\frac{\mu_0}{2F}{\bf M}^2+\frac{\mu_0\omega^2_0}{2F}\left(\int{\bf
M}dt \right)^2\right]\non\\
&&\hspace{3mm}+\frac{\gamma\mu_0}{F}{\bf M}^2.\label{magfinal}
\eqn The magnetic energy density is thus written as \beq
W_b=\frac{\mu_0{\bf H}^2}{2}+\mu_0{\bf H}\cdot{\bf
M}+\frac{\mu_0{\bf M}^2}{2F}+\frac{\mu_0\omega^2_0\left(\int{\bf
M} dt\right)^2}{2F}.\label{mage}\eeq The $\gamma\mu_0{\bf M}^2/F$
term in Eq.(\ref{magfinal}) represents the magnetic part of the
power loss $P_{loss}$ caused by the Joule heat in the split-ring
resonators, also consistent with Eq.(\ref{powerform})

Using Eq.(\ref{M}) once more, the magnetic energy density can be
rewritten as \bqn W_b&=&\frac{\mu_0{\bf H}^2}{2}+\mu_0{\bf
H}\cdot{\bf M}+\frac{\mu_0{\bf
M}^2}{2F}\non\\&&+\frac{\mu_0}{2\omega^2_0F}\left(\dot{\bf
M}+F\dot{\bf H}+\gamma{\bf M} \right)^2 \label{Wb0}\eqn or
expressed alternatively as \bqn W_b&=&\frac{\mu_0(1-F)}{2}{\bf
H}^2+\frac{\mu_0}{2\omega^2_0F}\left[\left(\dot{\bf M}+F\dot{\bf
H}+\gamma{\bf
M}\right)^2\right.\non\\&&\left.+\omega^2_0\left({\bf M}+F{\bf
H}\right)^2\right].\label{Wb}\eqn Note that this final form of
magnetic energy density is similar to the Eq.(15) of
\cite{Boardman}, but they are different. Our derivation relies on
the knowledge of the correct form of the power loss, whereas the
derivation in \cite{Boardman} did not use this knowledge thus
ambiguity may arise as has been explained before.

The total power loss is given by \beq P_{loss}=\frac{\nu\dot{\bf
P}^2}{\omega^2_p\epsilon_0}+\frac{\gamma\mu_0{\bf
M}^2}{F}\label{ploss},\eeq which is indeed the expected form of
Eq.(\ref{powerform}), and different from the Eq.(16) of
\cite{Boardman}. Our power loss formula has clear physical
meaning, as has been explained before. On the other hand, the
Eq.(16) of \cite{Boardman} has no such clear and convincing
meaning, thus we believe it need to be modified. Note that the
difference between these two formulas (i.e., $P_{loss}-P_L$) is
given by \beq \Delta P_{loss}=\frac{\partial}{\partial
t}\left(\frac{\gamma^2\mu_0}{2\omega^2_0F}{\bf M}^2+\frac{\gamma
\mu_0}{F}{\bf M}\cdot{\int {\bf M}dt}\right). \eeq Since this is
simply a time derivative of a time varying quantity, it
contributes nothing to the time-averaged power loss if the field
is harmonic. The equivalence relation $\langle
P_{loss}\rangle=\langle P_L\rangle$ can be directly checked by
explicit calculation.

Now we consider the time averaged energy density for monochromatic
wave. Time averaging every term in Eq.(\ref{We}), we get the
electric energy density \beq \left\langle W_e\right\rangle
=\frac{\epsilon_0|{\bf
E}|^2}{4}\left(1+\frac{\omega^2_p}{\omega^2+\nu^2}\right). \eeq
Similarly, time averaging all the terms in Eq.(\ref{Wb}) and
adding them together, we get the magnetic energy density \beq
\left\langle W_b\right\rangle=\frac{\mu_0|{\bf
H}|^2}{4}\left[1+F\frac{\omega^2\left(3\omega^2_0-\omega^2\right)}
{\left(\omega^2_0-\omega^2\right)^2+\omega^2\gamma^2}\right].
\label{wbharmonic}\eeq We stress here that Eq.(\ref{wbharmonic})
is just the corrected result of the magnetic energy density
formula (31) in Ref.\cite{Tretyakov} after adding the mutual
induction energy term that mentioned in Section I.

\section{The ED-EC correspondence}

Referring to Ref.\cite{Chen}, we can now construct an EC model for
the SRR array. We will show that the magnetic energy density
formula (\ref{Wb}) can also be derived by virtue of this EC model.
The most important distinction between our following derivation
and those proposed by others is that we consider arbitrarily
varying physical quantities, whereas others considered the
restricted harmonic cases. The ED-EC correspondence further
confirms the correctness of our derived energy density and power
loss formulas.

To map the ED quantities to the corresponding EC ones, we adopt
the configuration sketched in figure 1. Accordingly, in one unit
cell, the SRRs are piled up in the $y$ direction to form an
SRR-stack, which can be viewed as a circular solenoid. The
y-spacing between two successive SRRs in one stack is $l$. These
SRR-stacks are periodically arranged at a square lattice of
lattice constant $a$. For one unit cell, in order to mimic the
magnetic field acting on the SRR-stack inside, we further
introduce an imagined cell-solenoid of square cross section,
wrapping around the ``unit cell tube". In one turn the coil line
of the cell-solenoid is assumed to spiral up $l$ in the $y$
direction. We will show in the following that by appropriately
defining the currents carried by the cell-solenoid and the SRR
stack inside the cell tube and the electromotive forces in them,
the ED-EC correspondence can indeed be established.

Now we define the physical quantities of the EC system. Since all
the vector quantities we considered are parallel to the $y$
direction, hereafter we treat them as scalar quantities. The
magnetization $M$ and the magnetic field $H$ (in the connected
region outside the SRR stacks) are given by \beq M=\frac{I\pi
r^2}{la^2}=F\frac{I}{l},\;\;H=H_{out}. \eeq Here $F=\pi r^2/a^2$
is the filling fraction of the SRR-solenoid in one unit cell.

The magnetic fields outside and inside an SRR-solenoid are \beq
H_{out}=H_0-\frac{\pi
r^2}{a^2}\frac{I}{l}=H_0-F\frac{I}{l}=H_0-M,\eeq \beq
H_{in}=H_{out}+\frac{I}{l}=H+\frac{M}{F}=\frac{M+FH}{F},\eeq
respectively. Here $H_0$ represents the incident magnetic field.

The self inductances per turn of the cell-solenoid $L_0$ and of
the SRR-solenoid $L$ as well as the mutual inductance ${\cal M}$
between them are given by\beq
L_0=\mu_0\frac{a^2}{l},\;\;L=\mu_0\frac{\pi r^2}{l}=FL_0,\;\;{\cal
M}=FL_0=L.\eeq

The currents flowing in the cell-solenoid ($I_0$) and in the
SRR-solenoid ($I$) per turn are \beq
I_0=Hl=H_0l-FI,\;\;\;\;I=\frac{Ml}{F}.\eeq

The ``pure magnetic energy" stored in one slice of the unit cell
tube of thickness $l$, without taking into account the energy
stored in the interior capacitor of the SRR, can be calculated:
\bqn &&\frac
{1}{2}L_0I^2_0+\frac{1}{2}LI^2+{\cal M}I_0I\non\\
&=&(a^2l)\left[\frac{\mu_0H^2}{2}+\frac{\mu_0M^2}{2F}+\mu_0HM\right]\non\\
&&=(a^2l)\left[\frac{\mu_0(1-F)H^2}{2}+\frac{\mu_0(M+FH)^2}{2F}
\right]\non\\&&=(a^2l)\left[
(1-F)\frac{\mu_0}{2}H^2_{out}+F\frac{\mu_0}{2}H^2_{in}\right].\eqn
The physical meaning of the last form is obvious since the volume
fractions are $F$ and $1-F$, respectively.

The charge $q$ and the corresponding energy stored in the interior
capacitor of the SRR are
\bqn q &=& \int I dt=\frac{l}{F}\int M dt\non\\
&=&-\frac{l}{F\omega^2_0}\left(\dot{M}+F\dot{H}+\gamma
M\right)\eqn and \beq
\frac{q^2}{2C}=(a^2l)\frac{\mu_0}{2\omega^2_0F}\left(\dot{M}+F\dot{H}+\gamma
M\right)^2, \eeq  respectively. The Joule heat generating in the
SRR can also be evaluated: \beq RI^2=(a^2l)\frac{\gamma
\mu_0M^2}{F}.\eeq Here we have used the defining relations
\cite{Tretyakov}: \beq \frac{1}{C}=\omega^2_0L,\;\;\;\;\;R=\gamma
L.\eeq From these results we conclude that the ED and EC
approaches are indeed equivalent. Besides, the physical meaning of
each term appearing in Eq.(\ref{Wb}) becomes very clear now.

\section{Conclusion}

In this paper, we review the energy density formulas obtained in
\cite{Tretyakov} and \cite{Boardman} and analyze the apparent
contradictions between the equivalent circuit (EC) and
electrodynamics (ED) approach. A small error in the magnetic
energy formula of \cite{Tretyakov} has been pointed out, and the
corrected EC energy formula was obtained. We show that energy
density of an arbitrarily varying electromagnetic wave in the
wire-SRR medium can be derived using either ED or EC approach, and
the results are consistent. Besides, our energy density formula
reduces to Landau's classical formula in the lossless limit. This
investigation reveals that the ED and EC approaches are equivalent
if the correct expression of power loss is known.

{\it Note added.} One reviewer of this paper pointed out that in
another publication of Prof. Tretyakov's group \cite{Tretyakov1},
the same kind of EC approach as they used in \cite{Tretyakov} were
used again to derive the field energy density. After carefully
checking every steps, we found their derivation of magnetic energy
formulas (Eq.(28) and Eq.(33)) was based on the Eq.(20), which
means the mutual inductance contribution was still omitted. Thus
the results in \cite{Tretyakov1} should also be corrected.
Finally, we must stress that any energy density formula for an
effective medium can only be used in the frequency range where the
effective theory is accurate enough, and one should not expect the
formula to give reliable result beyond this frequency range.

\section*{ACKNOWLEDGEMENT} The author gratefully acknowledge
financial support from National Science Council (Grant No. NSC
95-2221-E-008-114-MY3) of Republic of China, Taiwan.

%%%%%%%%%%%%%%%%%%%%%%%%%%%%%%%%%%%%%%%%%%%%%%%%%%%%%%%%%%%%%%%%%%%%%%%%%%%%%

\begin{figure}
    \includegraphics[width=7.5in]{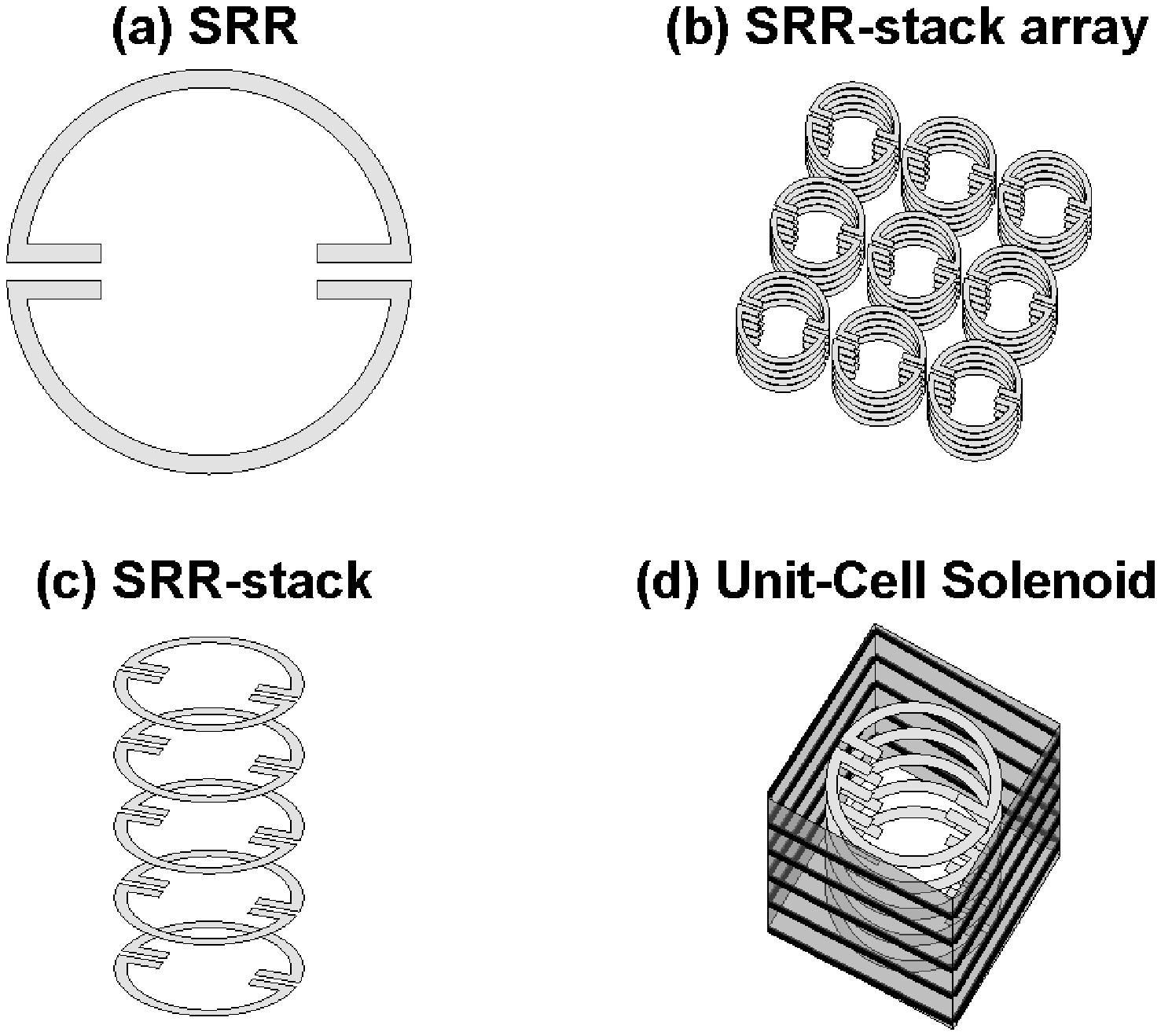}
    \caption{(a) SRR and unit cell. (b) SRR-stack array.
    (c) A SRR-stack as a solenoid. The $y$-spacing of two SRRs in a stack
    is $l$. (d) The ``unit cell tube" and the cell-solenoid around
    it.}
\end{figure}

\end{document}